\documentclass[a4paper,11pt]{article}
\usepackage{pos}
\usepackage{cleveref}
\title{Update of HPQCD $B_c\to J/\psi$ Form Factors}

\author*[a]{Judd Harrison}

\affiliation[a]{School of Physics and Astronomy,
  University of Glasgow, Glasgow, G12 8QQ, UK}

\emailAdd{judd.harrison@glasgow.ac.uk}

\abstract{We present an update of our lattice QCD determination of the $B_c\to J/\psi$ vector and axial-vector form factors, including new results for the tensor form factors. We use the Highly Improved Staggered Quark action for all valence quarks, together with the second generation MILC $n_f=2+1+1$ HISQ gluon field configurations. This calculation includes two additional ensembles, one with physically light up and down quarks and $a\approx 0.06 \mathrm{fm}$ and one with $a\approx 0.03\mathrm{fm}$ on which we are able to reach the physical bottom quark mass. Our calculation uses nonperturbatively renormalised current operators and covers the full kinematical range of the decay. Our physical-continuum extrapolation utilises the full dispersive parameterisation for $B_c\to J/\psi$.}

\FullConference{The 41st International Symposium on Lattice Field Theory (LATTICE2024)\\
 28 July - 3 August 2024\\
Liverpool, UK\\}

\begin{document}
\maketitle

\section{Introduction}

In recent years, lattice QCD studies of semileptonic $B$-meson decays have progressed significantly, with many new results away from zero recoil for decays to both vector and pseudoscalar final states~\cite{EuanBsDs,Cooper:2020wnj,Harrison:2020gvo,Harrison:2021tol,FermilabLattice:2021cdg,Cooper:2021bkt,Colquhoun:2022atw,Parrott:2022rgu,Flynn:2023nhi,Harrison:2023dzh,Aoki:2023qpa}. For $B\to D^{*}\ell\overline{\nu}$ decay, mediated by the Standard Model~(SM) charged-current interaction $b\to cW^-$, the new calculations from Fermilab-MILC, HPQCD and JLQCD have confirmed the long-standing tension between inclusive and exclusive determinations of $V_{cb}$~\cite{FlavourLatticeAveragingGroupFLAG:2024oxs}. While the form factors from these calculations can be fit with reasonable agreement, some disagreement is seen at the level of $1-2\sigma$ between predictions for observables such as $R(D^*)=\Gamma(B\to D^{*}\tau\overline{\nu}_\tau)/\Gamma(B\to D^{*}\mu\overline{\nu}_\mu)$ made using each calculation~\cite{Bordone:2024weh}. 

The Fermilab-MILC and JLQCD calculations generated lattice data only in the first $\approx 1/3$ and $\approx 1/4$ of the physical range of $w=v_{D^*}\cdot v_B$ respectively. The HPQCD calculation generated data across the full $w$-range only on the finest ensemble, and with limited statistical precision, resulting in greater uncertainties close to $w_\mathrm{max}$. To overcome these limitations, a BGL parameterisation~\cite{Boyd:1997kz} is typically used to fit and extrapolate synthetic datapoints for the form factors across the full kinematical region, so that predictions for integrated observables can be made. A relatively small under-estimation of uncertainties in the low-recoil region, particularly in the slope of the form factors, therefore has the potential to lead to a more significant underestimation of uncertainties in the high-recoil region, and thus in integrated observables such as $R(D^*)$. Moreover, since synthetic form factor data at different values of $w$ typically exhibit large correlations, systematic uncertainties related to higher-order kinematical depdendence may be mistakenly deemed insignificant if compared to the total uncertainty. 

The choice of kinematical parameterisation and truncation order used to perform the chiral continuum extrapolation is one potential origin for such underestimated uncertainties. For $B\to D^*\ell\nu$, expansions in powers of $(w-1)$ were used~\cite{FermilabLattice:2021cdg,Colquhoun:2022atw,Harrison:2023dzh}. Fermilab MILC and JLQCD fit the coefficients of $(w-1)^n$ in this expansion directly, including terms up to $(w-1)^2$, while HPQCD used terms up to $(w-1)^{10}$, with coefficients expressed as functions of the continuum BGL expansion parameters, including terms up to $z^4$. \cite{Harrison:2023dzh} also included a correlated fit of the $B_s\to D_s^*$ form factors, related to $B\to D^*$ by chiral parameters including polynomial terms in $M_\pi^2/\Lambda_\chi^2$, as well as the $SU(3)$ rooted-staggered chiral logarithms computed using perturbation theory. This update found differences of order $1-2\sigma$ for $h_{A_2}$ and $h_{A_3}$ relative to the previous $B_s\to D_s^*$ calculation~\cite{Harrison:2021tol}, due to the use of a BGL-like $z$-expansion omitting the outer functions to perform the chiral-continuum extrapolation in the earlier work.

As well as providing many important tests of SM lepton flavour universality~\cite{PhysRevLett.120.121801,CMS:2024uyo}, the related $B_c\to J/\psi\ell\nu$ decay provides an ideal testing ground to investigate the effects of the choice of parameterisation scheme on form factors and observables. The replacement of the light spectator quark with a heavier charm quark relative to the $B\to D^*$ case leads to improved statistical precision, enabling good resolution of the noisier $h_{A_2}$ and $h_{A_3}$ form factors (equivalently $A_{12}$ in the QCD basis), as well as reduced chiral effects due to differences in light quark masses. In this work, we provide an update of the HPQCD $B_c\to J/\psi$ form factors~\cite{Harrison:2020gvo}. In addition to methodological improvements and the calculation of the tensor form factors, this work includes the addition of an ensemble with physical light quarks and $a\approx 0.06 \mathrm{fm}$ and an ensemble with heavier-than-physical light quarks and $a\approx 0.03\mathrm{fm}$ on which we are able to simulate the physical $b$ quark with $am_b\approx 0.625$. This calculation includes two ensembles covering the full kinematical range with high statistical precision.

\section{Lattice Calculation}

\begin{table}
\caption{Details of the MILC $n_f=2+1+1$ HISQ gluon field configurations used in our calculation \cite{PhysRevD.87.054505,PhysRevD.82.074501}. The Wilson flow parameter~\cite{Borsanyi:2012zs}, $w_0$, used to fix the lattice spacing given is in column 3. We use the physical value of $w_0=0.1715(9)\mathrm{fm}$ determined in \cite{PhysRevD.88.074504}. The values of $w_0/a$, 
which are used together with $w_0$ to compute $a$ were taken from \cite{PhysRevD.96.034516,PhysRevD.91.054508,Hatton:2020qhk}. $n_\mathrm{cfg}$ is the number of configurations that we use here and $n_t$ is the number of time sources used on each configuration. $am_{l0}$, $am_{s0}$ and $am_{c0}$ are the masses of the sea up/down, strange and charm quarks in lattice units. We also include the approximate mass of the Goldstone pion, computed in~\cite{Bazavov:2017lyh}.\label{tab:gaugeinfo}}
\begin{tabular}{c c c c c c c c c c}\hline
 Set &$a$ &$w_0/a$& $L_x\times L_t$ &$am_{l0}$&$am_{s0}$& $am_{c0}$ & $M_\pi$ &$n_\mathrm{cfg}\times n_t $ \\ 
  & $(\mathrm{fm})$& & &&&  & $(\mathrm{MeV})$ & \\ \hline
1 & $0.0902$  & 1.9006(20) & $32\times 96 $    &$0.0074$ &$0.037$  & $0.440$ & $316$ & $1000\times 16$\\
2 & $0.0592$  & 2.896(6)   & $48\times 144  $    &$0.0048$ &$0.024$  & $0.286$ & $329$ & $500\times 4$\\
3 & $0.0441$  & 3.892(12)  &$ 64\times 192  $    &$0.00316$ &$0.0158$  & $0.188$ & $315$ & $375\times 4$\\
4 & $0.0327$  & 5.243(16)  &$ 96\times 288  $    &$0.00223$ &$0.01115$  & $0.1316$ & $309$ & $100\times 4$\\
5 & $0.0879$  & 1.9518(7)  &$ 64\times 96  $    &$0.0012$ &$0.0363$  & $0.432$ & $129$ & $600\times 8$\\
6 & $0.0568$  & 3.0170(23) & $96\times 192  $    &$0.0008$ &$0.0219$  & $0.2585$ & $135$ & $100\times 4$\\\hline
\end{tabular}
\end{table}

We generate correlation functions for $H_c$ and $J/\psi$ meson operators, including two- and three-point functions, using HISQ $h$ and $c$ valence quarks on the second generation MILC $n_f=2+1+1$ HISQ gluon field configurations listed in~\cref{tab:gaugeinfo}. The choices of staggered spin-taste operators, valence charm and heavy quark masses, and $J/\psi$ momentum, $\vec{p}'=(ak,ak,0)$, are identical to those used in~\cite{Harrison:2023dzh}. For the additional set with $a\approx 0.03\mathrm{fm}$, we use $am_h^\mathrm{val}=0.4,~0.625$, $am_c^\mathrm{val}=0.1316$ and $ak=0.0,~0.0500,~0.1501,~0.2501$.

We fit our lattice correlator data using the standard spectral decomposition for staggered correlation functions, implemented in the \textbf{corrfitter} python package~\cite{corrfitter}:
\begin{align}\label{twopointfit}
\langle  0|\bar{c}\gamma^\nu c(t) \big(\bar{c}\gamma^\nu c(0)\big)^\dagger| 0 \rangle =&\sum_{i}\Big((A^i_n)^2{e^{-tE^{i}_n}}\nonumber-(-1)^{t}(A^i_o)^2{e^{-tE^i_o}}\Big),\\
\langle  0|\big(\bar{h}\gamma^5 c(t)\big)^\dagger\bar{h}\gamma^5 c(0) | 0 \rangle =&\sum_{i}\Big((B^i_n)^2{e^{-tM^{i}_n}}-(-1)^{t}(B^i_o)^2{e^{-tM^i_o}}\Big)
\end{align}
and
\begin{align}\label{threepointfit}
\langle  0|\bar{c}\gamma^\nu c(T) ~ \bar{c}\Gamma h(t) ~ \bar{h}\gamma^5 c(0)| 0 \rangle &=\sum_{i,j}\Big({  A^i_n B^j_n J^{ij}_{nn} e^{-(T-t)E^{i}_n - tM^{j}_n} }\nonumber\\
+&{(-1)^{T+t}}  A^i_o B^j_n J^{ij}_{on} e^{-(T-t)E^i_o - tM^{j}_n} \nonumber\\
+&{(-1)^{t}}  A^i_n B^j_o J^{ij}_{no} e^{-(T-t)E^{i}_n - tM^j_o} \nonumber\\
+&{(-1)^{T}}  A^i_o B^j_o J^{ij}_{oo} e^{-(T-t)E^i_o - tM^j_o} \Big) .
\end{align}

The ground state parameters of these fits are related straightforwardly to the energies and matrix elements we require for computing form factors, giving
\begin{equation}
J^{00}_{nn(\nu,\Gamma)} = \sum_{\lambda}\frac{\epsilon^\nu(p',\lambda) \langle  J/\psi(p',\lambda ) |\bar{c}\Gamma h |H_c(p)\rangle}{\sqrt{2E_{J/\psi}2M_{H_c}\left(1+{\vec{p}}_{\nu}^{~\prime 2}/M_{J/\psi}^2\right)}}\label{relnorm}
\end{equation}
where ${\vec{p}}~'_{\nu}$ is the $\nu$ component of the $J/\psi$ spatial momentum, with $\nu$ corresponding to the Lorentz index of the $J/\psi$ vector interpolating operator. The matrix elements are related to the form factors in the HQET basis by
\begin{align}\label{formfactors}
\langle J/\psi|\bar{c}\gamma^5 h|\overline{{H_c}}\rangle             =& -\sqrt{M_{H_c}M_{J/\psi}}(\epsilon^*\cdot v)h_P,\nonumber\\
\langle J/\psi|\bar{c}\gamma^\mu h|\overline{{H_c}}\rangle           =&~ i\sqrt{M_{H_c}M_{J/\psi}}\varepsilon^{\mu\nu\alpha\beta}\epsilon^{*}_\nu v^\prime_\alpha v_\beta h_V,\nonumber\\
\langle J/\psi|\bar{c} \gamma^\mu \gamma^5 h|\overline{{H_c}}\rangle =& ~\sqrt{M_{H_c}M_{J/\psi}}\big[ h_{A_1}(w+1)\epsilon^{*\mu} -h_{A_2}(\epsilon^*\cdot v)v^\mu-h_{A_3}(\epsilon^*\cdot v)v^{\prime\mu} \big],\nonumber\\
\langle J/\psi|\bar{c}\sigma^{\mu\nu} h|\overline{{H_c}}\rangle      =& -\sqrt{M_{H_c}M_{J/\psi}}\varepsilon^{\mu\nu\alpha\beta}\big[ h_{T_1}\epsilon^*_\alpha(v+v^\prime)_\beta +h_{T_2}\epsilon^*_\alpha(v-v^\prime)_\beta +h_{T_3}(\epsilon^*\cdot v)v_\alpha v^\prime_\beta \big].
\end{align}
In this analysis, we will use the QCD basis in which the dispersive parameterisation is formulated~\cite{Gubernari:2023puw}. This basis is related to the HQET basis by
\begin{align}\label{formfactorsQCD}
V   = h_V\frac{1+r}{2 \sqrt{r}},   ~~~~A_0 = \frac{1}{2 \sqrt{r}}&\left(h_{A_1}(1+w)+h_{A_2}(rw-1)+h_{A_3}(r-w)\right),\nonumber\\
A_1 = \frac{\sqrt{r}}{1+r}h_{A_1}  (1+w),   ~~~&~A_2 = \frac{1+r}{2 \sqrt{r}}\left(h_{A_2} r+h_{A_3}\right),\nonumber\\
T_1 = -\frac{1}{2\sqrt{r}}\left(  (1-r)h_{T_2} - (1+r)h_{T_1}  \right),   ~~~&~T_2 = \frac{1}{2\sqrt{r}}\left( \frac{2r(w+1)h_{T_1}}{1+r} + \frac{2r(w-1)h_{T_2}}{1-r}  \right)\nonumber\\
T_3 = \frac{1}{2\sqrt{r}}\big(  (1-r)h_{T_1} - &(1+r)h_{T_2} + (1+r^2)h_{T_3}  \big).
\end{align}
it is conventional to define the related form factors
\begin{align}\label{helicityffsA12T23}
    A_{12} &= \frac{(M_{H_c} + M_{J/\psi})^2 (M_{H_c}^2 - M_{J/\psi}^2 - q^2)  A_1 - \lambda A_2}{16M_{H_c} M_{J/\psi}^2 (M_{H_c} + M_{J/\psi})}, \nonumber\\
    T_{23} &= \frac{(M_{H_c}^2 - M_{J/\psi}^2) (M_{H_c}^2 + 3 M_{J/\psi}^2 - q^2)  T_2 -  \lambda T_3}{8 M_{H_c} M_{J/\psi}^2 (M_{H_c} - M_{J/\psi})},
\end{align}
which diagonalise the bounds, where $\lambda=(M_{H_c}^2 - M_{J/\psi}^2 -q^2)^2-4M_{J/\psi}^2q^2$. 

The form factors must also satisfy kinematical constraints at $q^2=0$,
\begin{align}\label{kinz}
A_0(q^2=0) =& \frac{M_{H_c}+M_{J/\psi}}{2M_{J/\psi}}A_1(q^2=0)-\frac{M_{H_c}-M_{J/\psi}}{2M_{J/\psi}}A_2(q^2=0)\nonumber\\
T_1(q^2=0) =& T_2(q^2=0)\nonumber\\
\end{align}
and at $q^2_\mathrm{max}=(M_{H_c}-M_{J/\psi})^2$,
\begin{align}\label{kinmax}
A_{12}(q^2=q^2_\mathrm{max}) =& \frac{\left(M_{H_c}+M_{J/\psi}\right)\left(M_{H_c}^2-M_{J/\psi}^2-q^2_\mathrm{max}\right)}{16M_{H_c}M_{J/\psi}^2}A_1(q^2=q^2_\mathrm{max})\nonumber\\
T_{23}(q^2=q^2_\mathrm{max}) =& \frac{\left(M_{H_c}+M_{J/\psi}\right)\left(M_{H_c}^2+3M_{J/\psi}^2-q^2_\mathrm{max}\right)}{ 8M_{H_c}M_{J/\psi}^2}T_2(q^2=q^2_\mathrm{max}).
\end{align}
\subsection{Current Renormalisation}

In general, the lattice operators we use must be related to their continuum counterparts by renormalisation. Since we determine the pseudoscalar form factor via the partially conserved axial current relation, no renormalisation factor is required. For the vector current, the renormalisation factors, $Z_V$, were computed in the RI-SMOM scheme in~\cite{Hatton:2019gha} and~\cite{Hatton:2020qhk}. We also use these for the axial-vector current, using the chiral symmetry of the HISQ action~\cite{Sharpe:1993ur} together with the absence of condensate contamination in $Z_V$~\cite{Hatton:2019gha}. The tensor renormalisation factors, $Z_T$, were computed using an intermediate RI-SMOM scheme, matched to $\overline{\mathrm{MS}}$ at a scale $\mu=2\mathrm{GeV}$. We run these to $\mu=\overline{m}_h(\overline{m}_h)$ using the 3-loop anomalous dimension~\cite{Gracey:2000am} before multiplying our tensor matrix elements. For simplicity, we will leave the renormalisation factors implicit in the remaining discussion.

\subsection{Physical Continuum Extrapolation}
Following \cite{Gubernari:2023puw}, we use orthonormal polynomials on the unit circle, 
\begin{align}
\int_{-\alpha(t_X)}^{\alpha(t_X)} d\theta p_n(e^{i\theta})p_m(e^{-i\theta}) = \delta_{nm},
\end{align} 
where $\alpha(t_X) = \arg z(t_X,t_+,t_0)$, with $t_+=(M_H+M_{D^*})^2$ the two particle production threshold for the $\bar{c}\Gamma h$ current and $t_X=(M_{H_c}+M_{J/\psi})^2$ the start of the physical region for $H_cJ/\psi$ production. 

\begin{figure}
\centering
\includegraphics[scale=0.16]{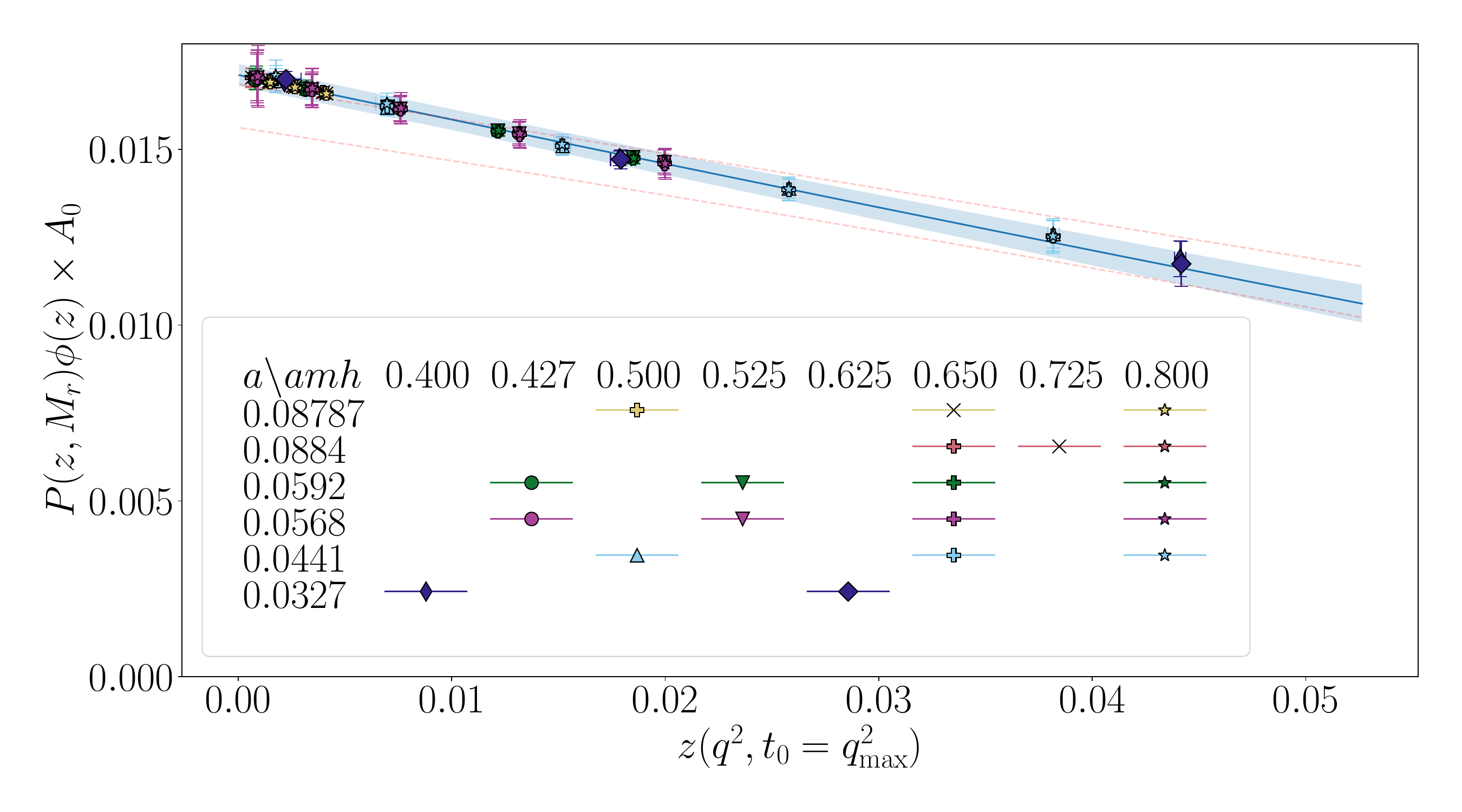}\includegraphics[scale=0.16]{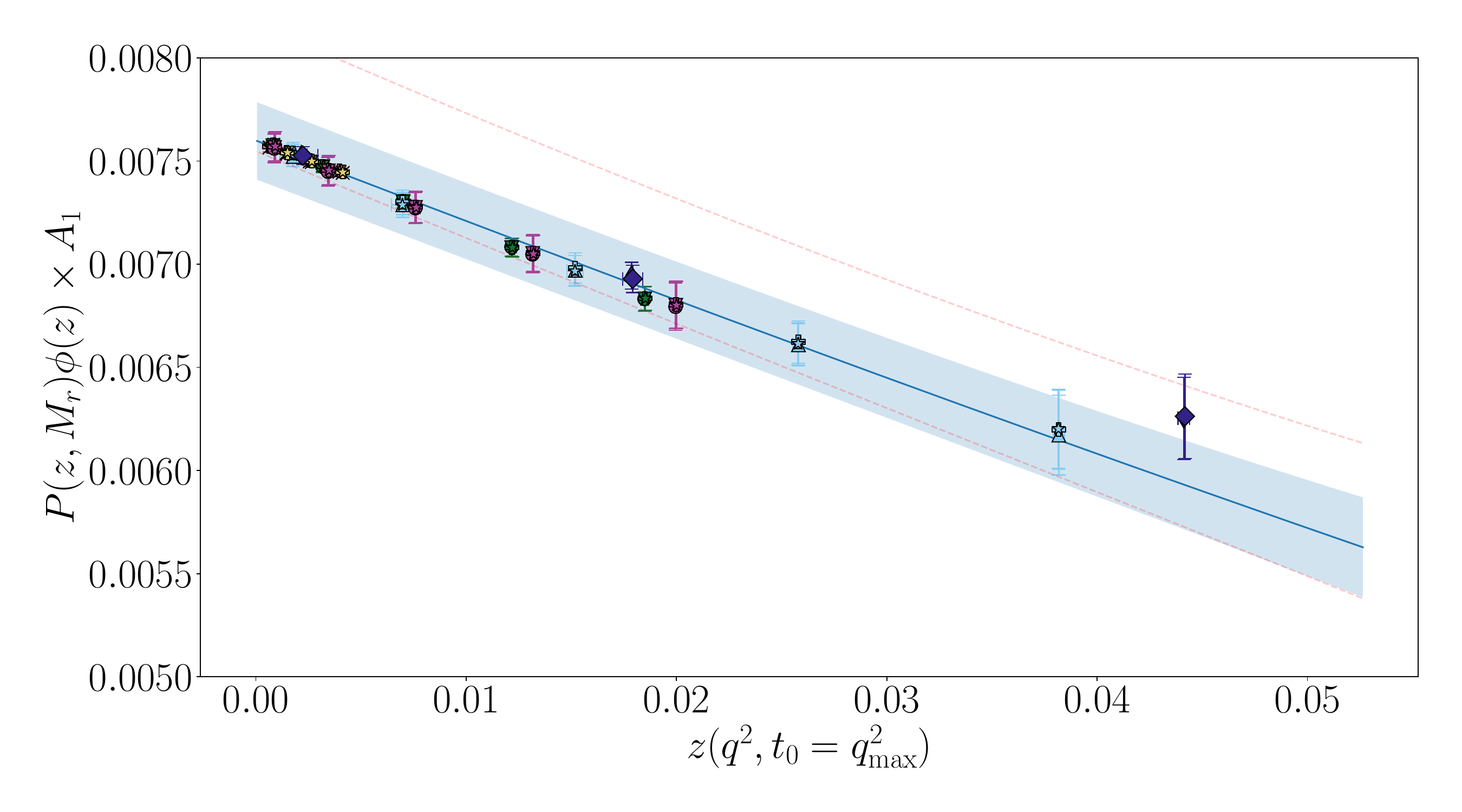}\\
\includegraphics[scale=0.16]{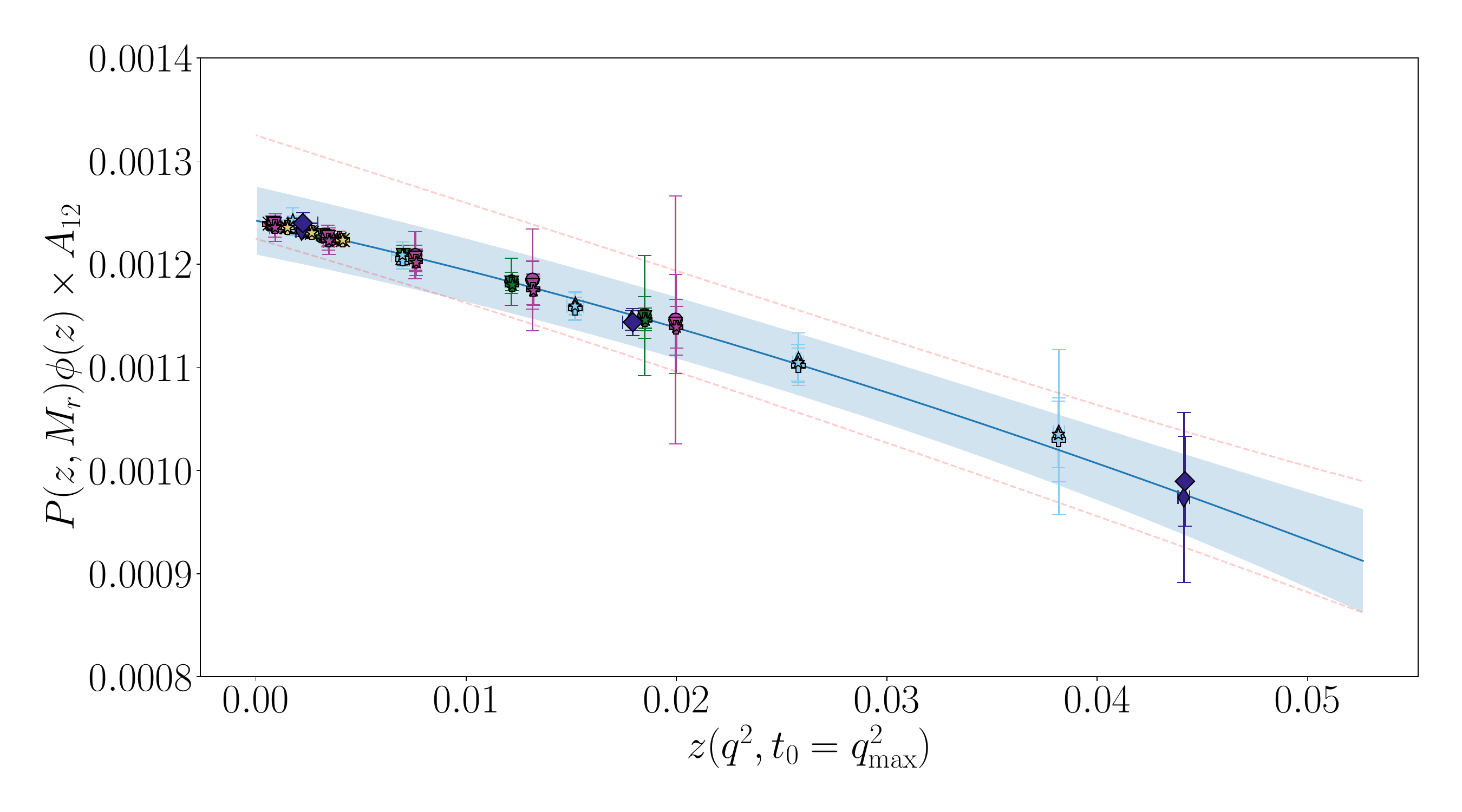}\includegraphics[scale=0.16]{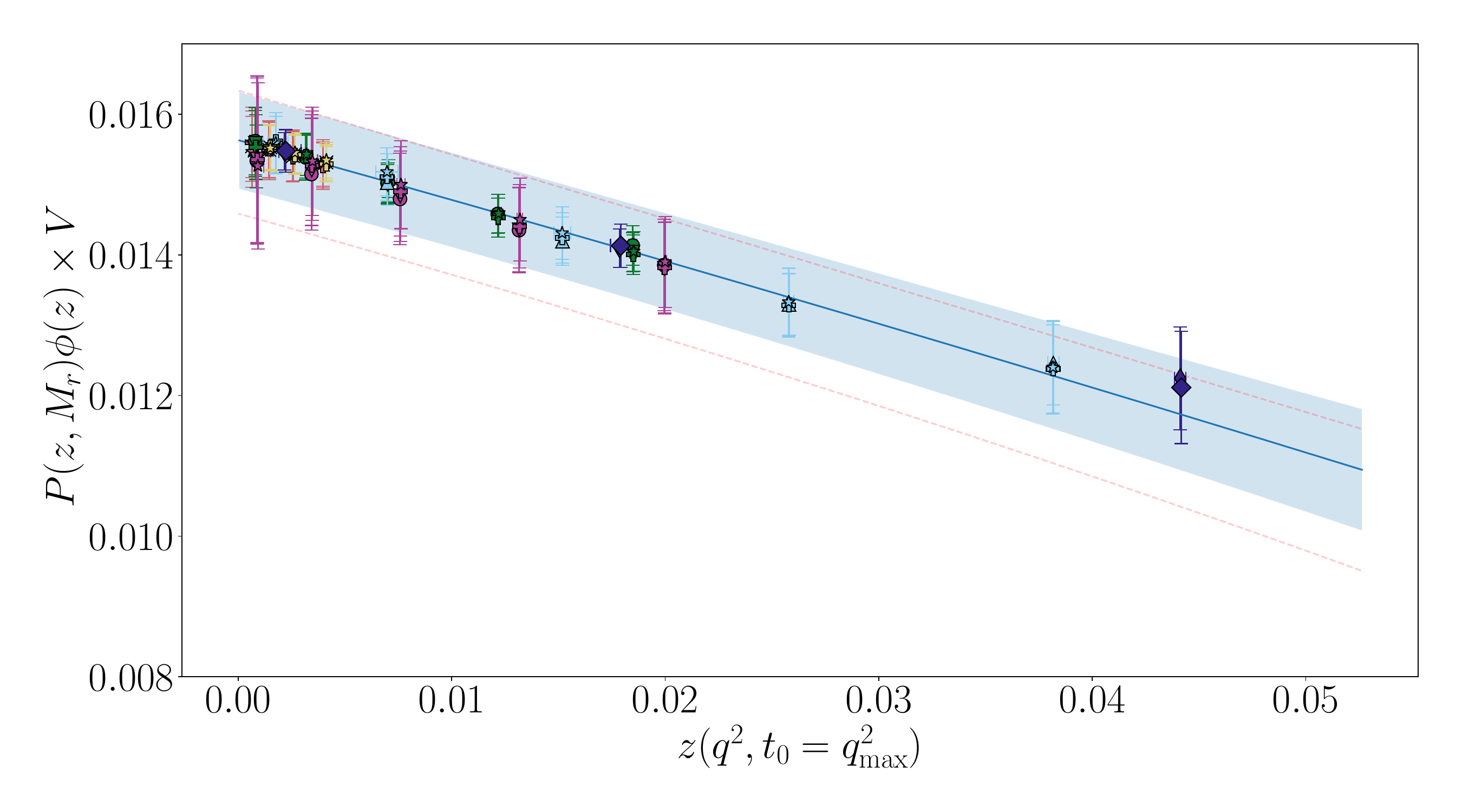}\\
\includegraphics[scale=0.16]{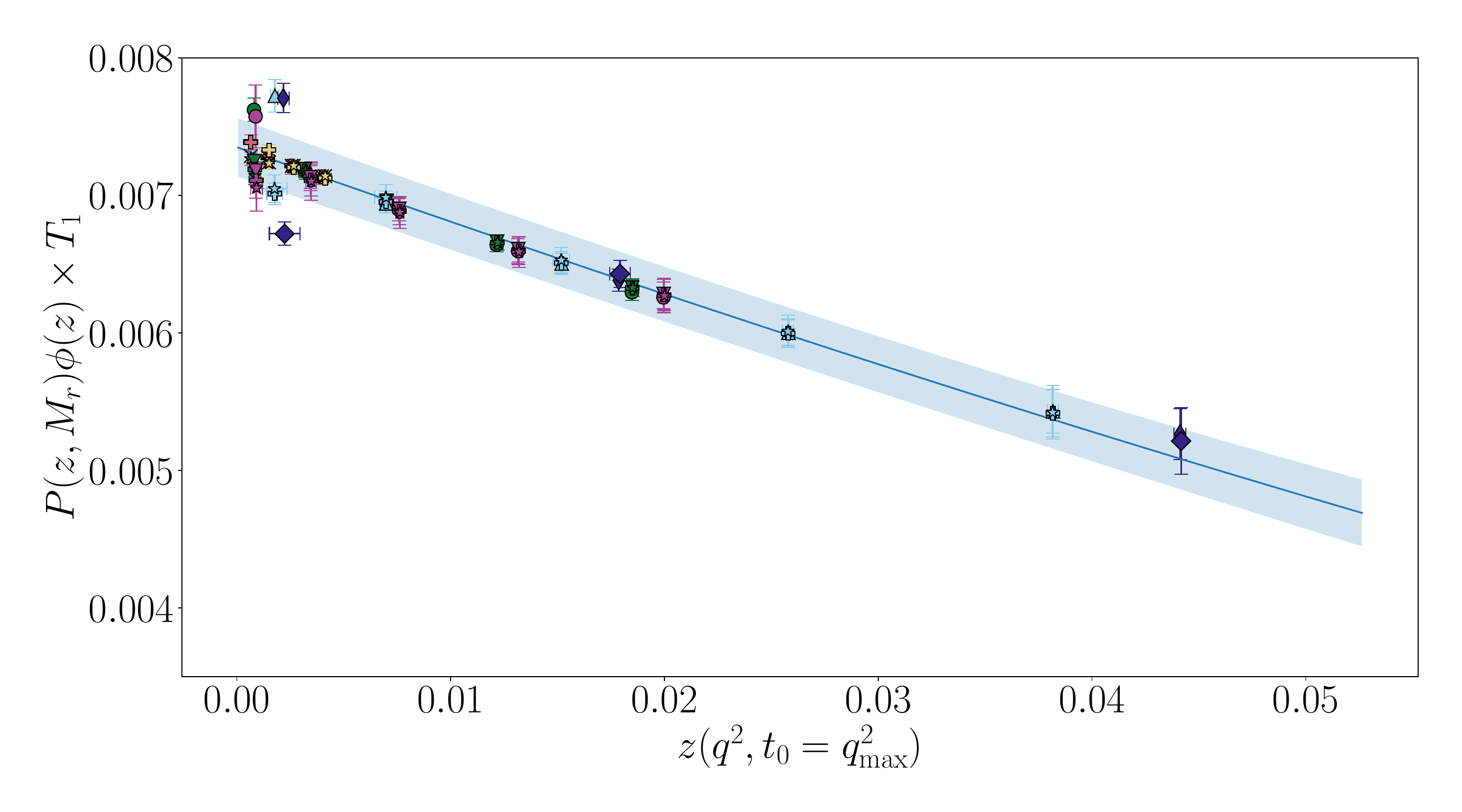}\includegraphics[scale=0.16]{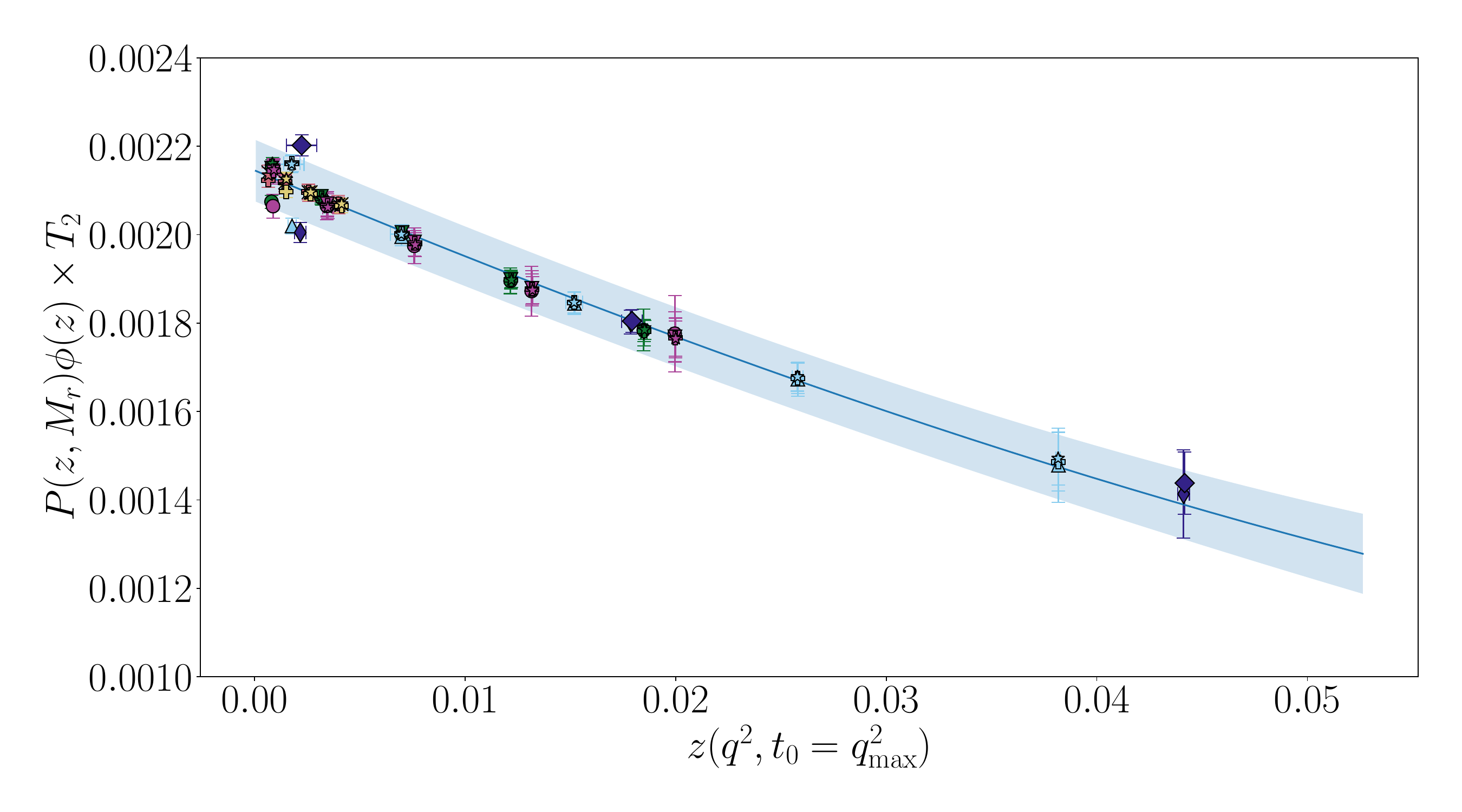}\\
\includegraphics[scale=0.16]{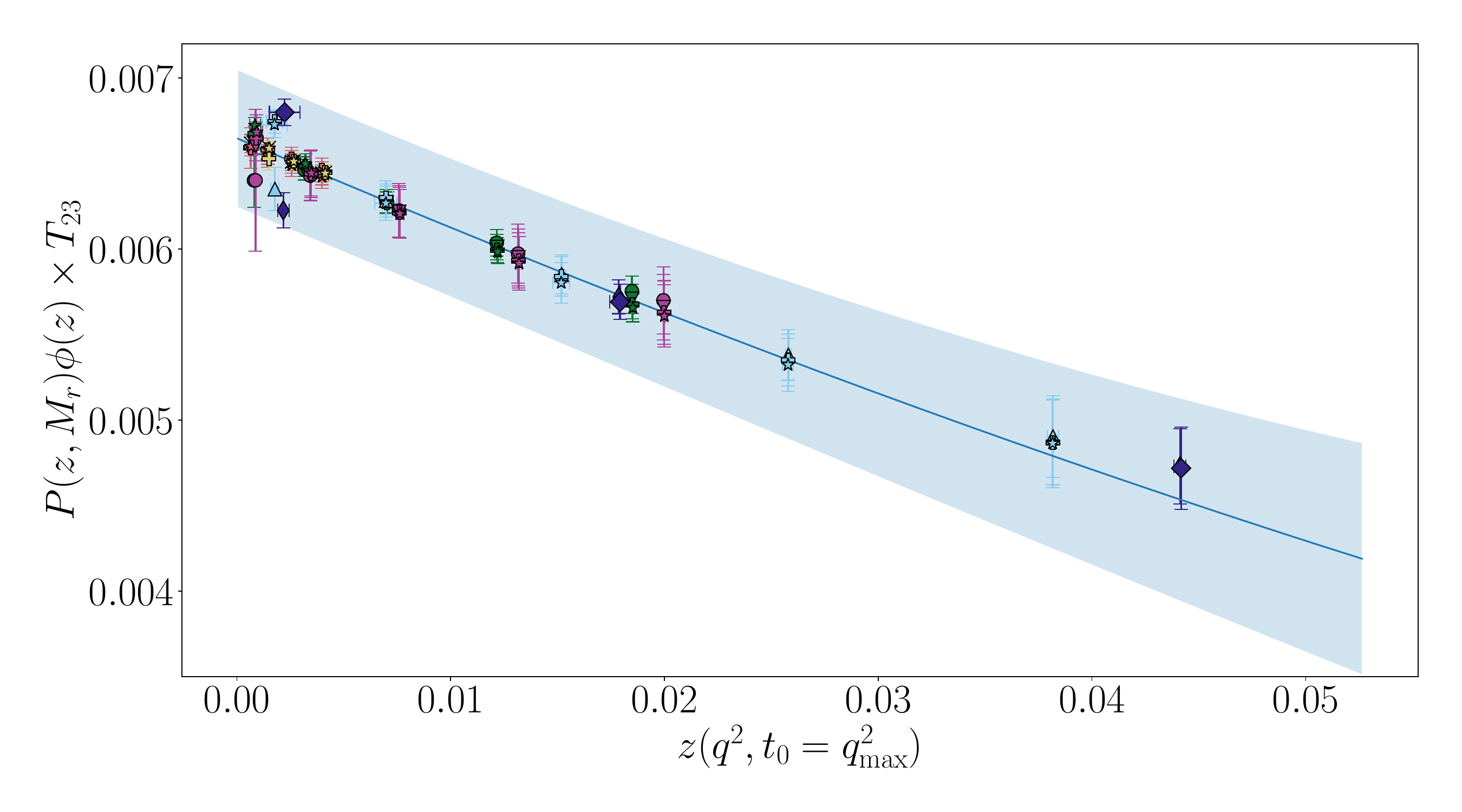}
\caption{\label{ffplots}Lattice data for all SM and tensor form factors together with the result of our physical continuum extrapolation described above, shown as the blue line and error band. The data points shown here have been corrected the posteriors, using~\cref{correction}. We also multiply the form factors by the Blaschke factors and outer functions to better demonstrate the polynomial dependence on $z$. The red dashed lines drawn for the SM form factors correspond to the $\pm 1\sigma$ confidence interval of our previous result~\cite{Harrison:2020gvo}.}
\end{figure}

We then use the dispersive parameterisation
\begin{align}\label{eq:ff}
F^Y(z) = &\frac{1}{P^Y(z,t_+,t_0)\phi^Y(z,t_+,t_0)}\sum_{n} a_n^Y p_n(z),
\end{align}
to describe the continuum form factors. Here, $P^Y(z(q^2,t_+,t_0)) = \prod_{i}z(q^2,t_+,M_{\mathrm{pole},i}^2)$ are Blaschke factors which depend on the masses, $M_{\mathrm{pole},i}$, of single particle $\bar{h}c$ states created by the current below the pair production threshold $t_+$. $\phi(z,t_+,t_0)$ are outer functions, defined in \cite{Gubernari:2023puw}, which are analytic on the open unit disk in $z$ and also depend on the susceptibilities, $\chi$. We use our recent lattice QCD calculation of the (pseudo)scalar, (axial-)vector and (axial-)tensor susceptibilities as a function of $u=m_c/m_h$, where, notably, some disagreement with $\mathcal{O}(\alpha_s)$ perturbation theory was seen for the (axial-)tensor susceptibility~\cite{Harrison:2024iad}. We use the simple heuristic forms $M_{\mathrm{pole},i} = M_{\mathrm{pole},i}^\mathrm{phys} + M_{H_c}^\mathrm{latt} - M_{B_c}^\mathrm{phys}$ and $M_H = M_B^\mathrm{phys} + M_{H_c}^\mathrm{latt} - M_{B_c}^\mathrm{phys}$ to approximate the heavy quark mass dependence of the meson masses we do not compute directly on the lattice. These forms ensure the correct values are reached in the physical continuum by setting $M_{H_c}^\mathrm{latt}=M_{B_c}^\mathrm{phys}$. The coefficients, $a_n^Y$, satisfy the bounds $\sum_{Y\to \Gamma}\sum_n |a_n^Y|^2$, where there is a bound for each current, and the sum over $Y$ runs over form factors which contribute in each case. In~\cref{eq:ff}, we include polynomials up to $\mathcal{O}(z^4)$ and impose the weaker bound $|a_n^Y|\leq 1$ by using uniform priors for $a_n^Y$ between -1 and 1. We include physical dependence on $m_h$ in the coefficients
\begin{align}
a^{Y}_n = \alpha^{Y}_n \times \Big(1+\sum_{j\neq 0}^3 b_n^{Y,j}&\Delta_{h}^{(j)} \Big)\mathcal{N}_n
\end{align}
where $\Delta_{h}^{(j= 0)}=1$ and
\begin{align}
\Delta_{h}^{(j\neq 0)}=\left(\frac{\Lambda}{2M_{H}}\right)^j-\left(\frac{\Lambda}{2M_{B}^\mathrm{phys}}\right)^j,
\end{align}
where we use $M_H$ as a proxy for $m_h$. We impose the kinematical constraints in~\cref{kinz,kinmax} for our continuum form factors at the physical point, and at one additional value of $M_{H_c}\approx (M_{B_c}+M_{\eta_c})/2$. We also include the mistuning term,
\begin{equation}
\mathcal{N}^{Y^{(s)}}_n = 1 + A^{Y}_n \delta_{m_c}^\mathrm{val}+ B^{Y}_n \delta_{m_c}^\mathrm{sea}+ C^{Y}_n \delta_{m_s}^\mathrm{sea}+D^{Y}_n \delta_{\chi}
\end{equation}
where
\begin{align}
\delta_{m_c}^\mathrm{val} &= (am_c^\mathrm{val}-am_c^\mathrm{tuned})/am_c^\mathrm{tuned},\nonumber\\
\delta_{m_c}^\mathrm{sea} &= (am_{c}^\mathrm{sea}-am_c^\mathrm{tuned})/am_c^\mathrm{tuned},\nonumber\\
\delta_{m_{s}}^\mathrm{sea} &= (am_{s}^\mathrm{sea} - am_{s}^\mathrm{tuned})/(10am_{s}^\mathrm{tuned}),\nonumber\\
\delta_\chi&=\left(\frac{M_{\pi}}{\Lambda_\chi}\right)^2-\left(\frac{M_{\pi}^\mathrm{phys}}{\Lambda_\chi}\right)^2,
\end{align}
where we use pion masses given in~\cite{Bazavov:2017lyh}. The tuned values of the charm and strange quark masses are given by
\begin{equation}
am_c^\mathrm{tuned} = am_c^\mathrm{val}\frac{M_{J/\psi}^\mathrm{phys}}{M_{J/\psi}},
\end{equation}
and
\begin{equation}
am_s^\mathrm{tuned} = am_s^\mathrm{val}\left(\frac{M_{\eta_s}^\mathrm{phys}}{M_{\eta_s}}\right)^2
\end{equation}
where we use the values of $M_{\eta_s}$ and $am_s^\mathrm{val}$ given in~\cite{EuanBsDsstar}.

Following~\cite{Harrison:2023dzh}, we fit the matrix elements that we extract on the lattice directly, allowing for discretisation effects using an additive term capturing momentum and heavy quark mass dependent discretisation effects:

\begin{align}
J^{00}_{nn(\nu,\Gamma)} = J_\mathrm{phys}^{\nu,\Gamma} + &\sum_{j,n=0}^3\sum_{k+l\neq 0}^3 c_n^{(\nu,\Gamma),jkl}\Delta_{h}^{(j)} (w-1)^n \left(\frac{am_c^\mathrm{val}}{\pi}\right)^{2k}  \left(\frac{am_h^\mathrm{val}}{\pi}\right)^{2l}
\end{align}
where $J_\mathrm{phys}^{\nu,\Gamma}$ is computed from the form factors determined using~\cref{eq:ff} combined with~\cref{formfactors} and~\cref{relnorm,formfactorsQCD,helicityffsA12T23}. The mixed terms appearing with $(am_h/\pi)^2$ and $w-1$ capture $(ap/\pi)^2$ momentum dependent discretisation effects. The SM form factors, $V,~A_0,~A_1$ and $A_{12}$, and tensor form factors, $T_1$, $T_2$, and $T_{23}$, corresponding to the matrix elements extracted from our correlator fits are shown in~\cref{ffplots} together with the form factors extrapolated to the physical continuum using our fit function described above. We multiply the form factors and continuum result by $P^Y(z,t_+,t_0)\phi^Y(z,t_+,t_0)$ to illustrate the simple dependence on $z$. To better illustrate the consistency between our data and fit function, we also correct the data using the dependence on $a$, $am_h$ and $\Lambda_\mathrm{QCD}/M_H$ determined from our fit, so that what we plot is 
\begin{align}\label{correction}
F^\mathrm{corrected}(a,am_h,\Lambda_\mathrm{QCD}/M_{H},w)&=F^\mathrm{data}(a,am_h,\Lambda_\mathrm{QCD}/M_{H},w) \nonumber\\
&+ \Big[F^\mathrm{phys}(w)-F^\mathrm{fit}(a,am_h,\Lambda_\mathrm{QCD}/M_{H},w)\Big].
\end{align}
We see that our fit function captures the simple $z$ dependence of our data well. In~\cref{ffplots}, we also show the $\pm 1\sigma$ confidence interval of our previous results for the SM form factors, $V,~A_0,~A_1$ and $A_{12}$~\cite{Harrison:2020gvo} as red dashed lines. These show good agreement with our new, more precise results.

\section{Conclusions and outlook}
The calculation presented here includes additional lattice data for the $B_c\to J/\psi$ form factors on two new ensembles, one with physically light up and down quarks and $a\approx 0.06 \mathrm{fm}$ and one with $a\approx 0.03\mathrm{fm}$ on which we are able to reach the physical bottom quark mass. Utilising our recent lattice QCD calculation of the $\bar{h}c$ susceptibilities as a function of $u=m_c/m_h$, we use the full dispersive parameterisation including outer functions and Blaschke factors to extrapolate our data to the physical bottom quark mass in the chiral-continuum limit. Our results are found to be consistent with our previous calculation, though roughly a factor of 2 more precise. We also present new results for the $B_c\to J/\psi$ tensor form factors, which are vital to constraining new physics effects in this decay mode. Our calculation achieves excellent coverage of the kinematical range, with data on two ensembles spanning the full range. This will allow for a detailed study of systematic uncertainties associated with the choice of kinematical parameterisation used to perform the physical-continuum extrapolation.

\acknowledgments

We are grateful to the MILC Collaboration for the use
of their configurations and code. We thank C. T.H. Davies, C. Bouchard, D. van Dyk, M. Reboud, M. Jung and M. Bordone for useful discussions. Computing was done on
the Cambridge service for Data Driven Discovery (CSD3),
part of which is operated by the University of Cambridge
Research Computing on behalf of the DIRAC HPC Facility
of the Science and Technology Facilities Council (STFC).
The DIRAC component of CSD3 was funded by BEIS
capital funding via STFC capital Grants No. ST/P002307/1
and No. ST/R002452/1 and by STFC operations Grant
No. ST/R00689X/1. DiRAC is part of the national
e-infrastructure. We are grateful to the CSD3 support staff
for assistance. Funding for this work came from UK Science
and Technology Facilities Council Grants No. ST/L000466/1 and No. ST/P000746/1 and Engineering and Physical
Sciences Research Council Project No. EP/W005395/1.

\bibliographystyle{JHEP}
\bibliography{BcJpsiproc}

\end{document}